\documentclass[a4paper,twoside,10pt]{article}
\usepackage{graphicx,publaob,amsmath}
\usepackage{graphicx}

\def\be{\begin{equation}}
\def\ee{\end{equation}}
\def\bea{\begin{eqnarray}}
\def\eea{\end{eqnarray}}

\def\papertype{Contributed paper}
\begin{document}

% Title
\title{ GRAVITATIONAL MICROLENSING AND DARK MATTER PROBLEM:
RESULTS AND PERSPECTIVES}

% Authors
\authors{A.F. ZAKHAROV$^{1,2}$}

% Addresses and e-mails
\address{$^1$ Institute of Theoretical and Experimental Physics,
 117259,  B. Cheremushkinskaya, 25, Moscow, Russia
}
\Email{zakharov}{vitep1.itep}{ru}
\address{$^2$ Astro Space Centre of Lebedev Physics Institute, Moscow, Russia}

% Running titles
\markboth{ GRAVITATIONAL MICROLENSING AND DARK MATTER PROBLEM}{A.F. ZAKHAROV}

%\date{}

%\maketitle

\abstract{
Foundations of standard theory of microlensing are described,
namely we consider microlensing  stars in Galactic bulge, 
the Magellanic Clouds or other nearby galaxies.
We suppose that gravitational microlenses lie between an Earth observer and
these stars. 
Criteria of an identification of microlensing events are discussed.
We also consider such microlensing events which do not satisfy these
criteria (non-symmetrical light curves, chromatic effects,
polarization effects).
We describe  results of MACHO collaboration
observations towards the  Large Magellanic Cloud (LMC)
and the Galactic bulge in detail.
Results of EROS observations towards the LMC and OGLE observations towards
the Galactic bulge are also presented.
A comparison of the microlensing theory and observations is discussed
in full.}

%\section{Standard model}

%\subsection{Equation of motion of light rays}

A standard microlens model is based on a simple approximation of
a point mass for a gravitational microlens.

In the framework of general relativity using
 a weak gravitational field approximation the correct bending angle
is described by the following expression: 
\begin{eqnarray} 
\delta \varphi=-\frac{4GM_*}{c^2 p}.
%\#
\label{eqs5}
\end{eqnarray}
The derivation of the famous Einstein's formulae 
for the bending angle of light rays in gravitational field of a point mass
$M_*$ is practically in all monographs and textbooks 
on general relativity and gravity theory 
(see, for example books of Landau \& Lifshitz (1975), 
M\"oller (1972)).

In the framework of general relativity the light ray bend effect
was predicted by A. Einstein in 1915 
and  was firstly confirmed by Sir A. Eddington for observations
of light ray bend by the Solar gravitational field near its surface.
The angle is equal to $1.75''$, and Einstein prediction was
confirmed by observations.

%\subsection{Point lens equation}

\begin{figure}[tbh]
%\vspace{-40mm}
%\grpicture{figs2}
\begin{center}
\includegraphics[width=0.47\textwidth]{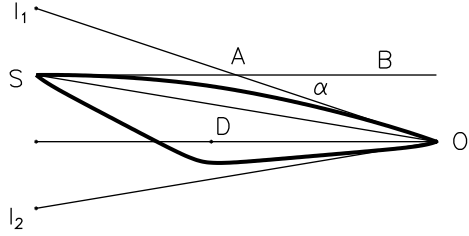}
\end{center}
\caption{Formation of images and light rays
bended by the gravitational field of a body.
}
\vspace{-3mm}
\label{figs2}
\end{figure}

Since a photon moves practically along straight lines far from
a gravitating body, we approximate the photon trajectory
by two straight lines which are intersected near the body $D$ 
(Fig. \ref{figs2}). 
The angle between the lines  $\alpha$ demonstrates
the photon bending in the gravitational field of the body $D$. 

Two rays of light, which lie in opposite sides of
the gravitating body, are deflected to the gravitating body. 
If a source $S$ lies far away from the body $D$ then the rays
begin to converge and intersect in some distant point $O$ (Fig. \ref{figs2}).
If we suppose that an observer is in the point $O$, he will see two
images ($I_1$, $I_2$) of one source $S$. 
Really that is gravitational lens effect.
In Fig. \ref{figs2} three physical bodies are shown,
namely the source $S$, the gravitating body $D$ 
and the observer $O$. Trajectories of light rays 
from $S$ to $O$  are shown by two bold solid lines.
We use also the following notations:
$D_{ds}$ is the distance from the source $S$ to the lens $D$; 
$D_{d}$ is the distance from  the lens $D$ to the observer $O$;
$D_{s}$ is the distance from the source $S$ to the observer $O$. 
We draw plane via the point $S$ and we suppose that the plane is
perpendicular to a light ray trajectory. 
The plane is called as a plane source.
Similarly, we draw the plane via the gravitational lens $D$.
The plane is called as a lens source.
We also use the following notations for the angles:
$\theta$ is the angle between a direction to the gravitating body 
$D$ and a direction to the source $S$,
$\theta_1$ is the angle between a direction to the gravitating body $D$ 
and   an apparent direction to the source $S$, 
$I_1$  $I_2$ are images (mirages) of the source. 

It can be seen from the figure that we have the following
expressions for the angles (Zakharov \& Sazhin (1998)):
\begin{eqnarray}
\alpha=\beta_1 +\beta, 
%\# \\
\label{eqs6}
\theta_1=\theta +\beta, 
%\# \\ 
\label{eqs7}
\beta_1 \cdot D_{ds}=\beta \cdot D_{d}, 
%\#
\label{eqs8}
\end{eqnarray}
where the angle $\beta$  is expressed in radians,  $D_{ds}\, , D_{d} $ 
are the distances from the source to the gravitational lens and
from the gravitational lens to the observer, respectively. 
From Eqns (\ref{eqs8})  
we obtain a quadratic equation  for the angle $\theta_1$
which determines apparent positions of images relatively
a direction toward a gravitational lens, 
\begin{eqnarray}
\theta_1^2 -\theta \theta_1 - \theta_0^2=0,
%\# 
\label{eqs9}
\end{eqnarray}
where $\theta$ is the angle between the direction
toward a gravitational lens  (GL)  and the true position of a distant
lensed source,  $\theta_0$ is the angular radius of the Einstein cone
which is defined as 
$$
\theta_0^2=\frac{4 G M}{c^2}\cdot \frac{D_{ds}}{(D_{ds} +D_{d})\cdot D_{d}}. 
%\#
$$ 
Eqn (\ref{eqs9}) is called the gravitational lens equation for
the case of a spherically symmetric point lens.
The equation has two real roots, namely
$$ 
\theta_1=\frac{1}{2}\theta +\frac{1}{2}\sqrt{\theta^2 +4\theta^2_0},
%\#
%$$
%
%
%$$
\quad
\theta_2=\frac{1}{2}\theta -\frac{1}{2}\sqrt{\theta^2 +4\theta^2_0},
%\#
$$
corresponding to two images of a source $S$. 
The angular distance between the images is equal 
to $\sqrt{\theta^2 +4\theta_0^2}$.

According to previous arguments we wrote (continuously) about two images.
However, these two images are formed not always. 
Really, we used the assumption that the sizes of a gravitating body $D$ 
are infinitesimal and the Eqn (\ref{eqs5}) is valid for any impact parameter.
Actually, if the impact parameter is smaller than the radius $R_D$ of
 a gravitating body $D$ or
$$
R_D>D_{d} \theta_2
%\#
$$
then the image $I_2$ disappears for an observer $O$ 
(the light ray moving along the trajectory with the impact parameter
is absorbed by a matter of a gravitational lens if it is non-transparent). 
Therefore, only one image of a source is formed for this case.
That is the reason why Earth's observer does not see two images
during a solar eclipse in spite of an existence of a set of stars
which lie near the line drawing via the solar center and the observer
(we recall that the angular solar size is about a half of an angular degree,
which is much greater than the Einstein's cone size of the Sun since
the distance between Earth's observer and the gravitational lens (Sun)
is equal to 1 astronomical unit).

It is necessary to note that according to the equivalence principle
two bodies with different masses fall with the same  acceleration
in a gravitational field.
Therefore, two photons having different frequencies
(different energies and thus different masses) are accelerated
identically in a gravitational field.
In other words, photons lying in different bands are bended
identically in a gravitational field of a body  $D$. 
This property is called the achromatism of the microlensing effect.
Possible violations of the property may be connected with
complicated structure of a source $S$, the violations will be discussed below.

The gravitational lens effect is a formation 
of several images instead of one. We have two images
for a point lens model (Schwarzschild lens model).
The angular distance between two images is about
angular size of so-called Einstein's cone.
The angular size of Einstein's cone is proportional to the lens mass
divided by the distance between a lens and an observer.
Therefore, if we consider a gravitational lens with typical
galactic mass and a typical galactic distance between a gravitational lens 
and an observer then the angular distance between images
will be about few angular seconds; 
if we suppose that a gravitational lens has a solar mass and
a distance between the 
lens and an observer is about several kiloparsecs then
an angular distance between images will be about angular millisecond.

If a separation angle is $\sim 1''$,
then one may observe two images in optical band 
although this problem is a complex  one,
but one cannot observe directly two images by Earth's
observer in the optical band if a separation angle is
$\sim 0.001''$. Therefore, the effect is observed
on changing of a luminosity of a source $S$.

If the source $S$ lies on the boundary of the Einstein cone 
($\theta (t)=\theta_0$), then we have  $A=1.34$. 
Note, that the total time of crossing of the Einstein cone is $T_0$, so
$$
T_0=2\frac{\sqrt{\theta_0^2-\theta^2_p}}{\Omega}.
%\#
$$
Sometimes the microlensing time is defined as a half of $T_0$. 
If we suppose that $D_{d}<D_{ds}$,  then
$$
T_0=3.5~ months \cdot \sqrt{\frac{M}{M_{\odot}} \frac{D_{d}}{10\, kpc}} 
\cdot \frac{300\, km/s}{v},
%\#
$$
where $v$ is the perpendicular component of a velocity
of a dark body.

We will give numerical estimations for parameters of the microlensing effect.
If the distance between a dark body and the Sun is equal to $\sim 10$~kpc, 
then the angular size of Einstein cone of the dark body with
a solar mass is equal to  $\sim 0.001''$ or the linear size of Einstein cone
is equal to about 10 astronomical units.
If we suppose that the perpendicular component of a velocity
of a dark body is equal to $\sim 300$ km/s (that is a typical
stellar velocity in Galaxy), then a typical time of 
crossing Einstein cone is about 3.5 months.  A luminosity of a source  $S$ 
is changed with the time.

For observations of ~extragalactic gravitational lens
a typical time for changes of light curve is very long
 ($\sim 10^{5}\!$ years) for its direct observations. 
Therefore, extragalactic gravitational lenses
are discovered and observed by resolving different optical components
(images) since typical angular distances between images
are about some angular seconds because of a great mass of a gravitational lens.
If a gravitational lens is a galaxy cluster then 
the angular distances between images may be about several minutes.
For an identification of gravitational lenses, 
observers compare typical features and spectra of different images.
It is clear that one cannot to resolve different components during microlensing
but it is possible to get and analyse a light curve in different
spectral bands.

%\section{Non-symmetrical light curves}

One of the basic criterion for microlensing event identification
is the symmetry of a light curve.
If we consider a spherically symmetric gravitational field of a 
lens, a point source and a short duration of microlensing event
then the statement about the symmetry of a light curve
will be a strong mathematical conclusion, but
if we consider a more complicated distribution 
of a gravitational field lens or an extensive
light source then some deviations of symmetric light curves
may be observed and (or) the microlensing effect 
may be chromatic (see details in Zakharov (1997); Zakharov \& Sazhin (1998)).

%\input rev6wof1.tex

%\section{Gravitational microlens observations}
%\subsection{Introduction}

For the first time a possibility to discover microlensing using
observations of star light curves was discussed in the paper by Byalko
(1969).
%\cite{bya69}.  
Systematic searches of dark matter using typical variations
of light curves of separate stars from millions observable stars
started after Paczynski's discussion (1986) of the halo dark matter discovery
using monitoring stars from Large Magellanic Cloud
(LMC). %\cite{pac86}.  
We remark that in the beginning of the nineties
new computer and technical possibilities providing
the storage and processing of huge volume of observational data
were appeared and it promoted at the rapid realization
of Paczynski's proposal.
Griest (1991) suggested to call the microlenses as MACHO 
(Massive Astrophysical Compact Halo Objects).  
Besides,  MACHO is the name of the project
of observations of the US-English-Australian collaboration
which  observed the LMC and Galactic bulge using 1.3 m telescope
of Mount Stromlo observatory in Australia.\footnote{MACHO stopped 
since end 1999.}
%Some information about the experiment is in the sites
%{\tt http://wwwmacho.mcmaster.ca/} and
% {\tt http://wwwmacho.anu.edu.au/}.
%Information about alert microlensing events from current
%observational data of MACHO collaboration
%in real time is in the site {\tt http://darkstar.astro.washington.edu/}.

The first papers about the microlensing discovery were published by
the MACHO collaboration (Alcock et al. (1993)) and the French collaboration EROS 
(Exp\'erience de Recherche d'Objets Sombres)
(Aubourg et al. (1993)).\footnote
{EROS experiment will stop in 2002 (Moniez, 2001)).}
%\cite{aub93}. 
%Some information about EROS experiment is in the sites
%{\tt http://www.lal.in2p3.fr/EROS/eros.html}.

First papers about the microlensing discovery
toward Galactic bulge were published by US-Polish collaboration (Optical
Gravitational Lens Experiment), which used
1 m telescope at Las Campanas Observatory. 
%Some information about the OGLE experiment is in the sites
%{\tt http://www.astrouw.edu.pl} and
%{\tt http://www.astro.princeton.edu/~ogle/}.  
%The results of the OGLE collaboration which include the photometry
%of OGLE microlensing event candidates, papers of the OGLE collaboration,
%as well as regularly updated status report      
%can be found over Internet from the host
%{\tt sirius.astrouw.edu.pl} (148.81.8.1) using "anonymous ftp"
%service.

%\subsection{Microlensing features}

The event corresponding to microlensing may be characterized
by the following main features, which allow to distinguish
the microlensing event and a stellar variability
(Roulet \& Mollerach (1997); Zakharov (1997)).  \\ 
\begin{itemize}

\item 
Since the microlensing events have a very small probability,
the events should never repeat for the same star.
The stellar variability is connected usually with
periodic (or quasi-periodic) events of the fixed star.
\item 
In the framework of a simple model of microlensing
when a point source is considered,
the microlensing effect must be achromatic
(deviations from achromaticity
for non-point source were considered, for example in the paper
by Bogdanov \& Cherepashchuk (1995)), but the proper change 
of luminosity star is connected usually
with the temperature changes and thus
the light curve depends on a colour.
\item 
The light curves of microlensing events are symmetric, but
the light curves of variable stars are usually
asymmetric (often they demonstrate
the rapid growth before the peak and the slow decrease 
after the peak of a luminosity).
\item
Observations of microlensing events are interpreted quite well
by the simple theoretical model, but some microlensing events
are interpreted by more complicated model
in which one can take into account that
a source (or a microlens) is a binary system,
a source has non-vanishing size, the parallax effect may take place.
\end{itemize}

\begin{figure}[!t]
%\vspace{7.5cm} 
%\special{psfile=fig8.ps voffset=-20 hoffset=46 vscale=30 hscale=30 angle=0} 
\begin{center}
\includegraphics[width=0.445\textwidth]{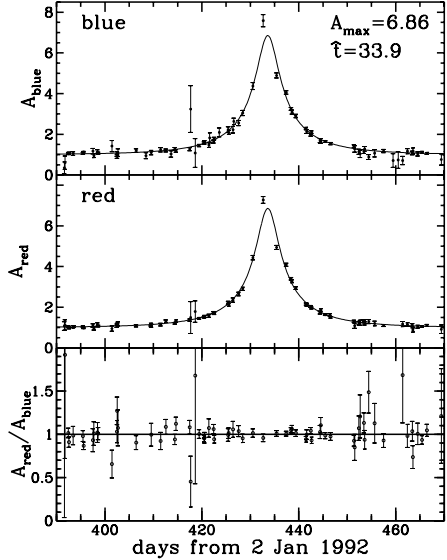}
\end{center}
\caption{The first microlensing event
which was detected by the MACHO collaboration
during microlensing searches towards LMC
(Alcock et al.~ (1993)).  } 
\label{gold} 
\end{figure}

The typical features of the light curve of the first
microlensing event observed by the MACHO collaboration
in the LMC  are shown in Fig.
\ref{gold}, where the light curves are shown for two spectral 
bands.\footnote{A more recent MACHO fit to the observed amplification
of this event gives $A_{\rm max}=7.2$.} 
The light curve (in two bands) is fitted by the simple model
well enough, but the ratio of luminosities for the bands is
shown in the lower panel of figure
(the ratio shape is adjusted with the event achromaticity).
However, one can note that near the maximal observable luminosity
the theoretical curve  fits the data of observations not very well.

Now one can carry out accurate testing  the achromaticity
and moreover the stability of the source spectrum
during a microlensing event with the Early Warning systems
implemented both by the MACHO  and OGLE 
collaborations. This allows  one to study the source properties
using large telescopes and to organize
intense follow-up studies of light curves
using telescope network around the globe.

In addition to the typical properties of individual microlensing events,
Roulet and Mollerach note that the population
of observed events should have the following statistical properties
(Roulet \& Mollerach (1996), Zakharov (1997)).

Unlike a star variability microlensing events should happen
with the same probability for any kind of star
therefore the distribution of microlensing events
should correspond to the distribution of observed stars
in the color-magnitude diagrams.\footnote{However,
Roulet and Mollerach (1997) noted that for observations in the bulge since
observed stars have non-negligible spread along the line of sight,
the optical depth is significantly
larger for the star lying behind the bulge,
thus the lensing probabilities should increase
for the fainter stars.}

The distribution of the maximal amplification factor $A_{\rm max}$ 
should correspond to a uniform distribution
of the variable  $u_{\rm min}$.

The distributions of the amplification $A_{\rm max}$ 
and the microlensing event time $T$ should be uncorrelated.

Since for the microlens searches one can monitor several 
million stars during several years, the ongoing searches
have focused on two targets:
a) stars in the Large and Small Magellanic Clouds (LMC and SMC)
which are the nearest galaxies having
lines of sight which go out of the Galactic
plane and well across the halo;
b) stars in the Galactic bulge which allow to test 
the distribution of lenses near to the Galactic plane.

%\subsection{Results and unsolved problems}

Let us cite well established results of microlensing searches
and discuss the questions for which we have now different answers
which do not contradict to the observational data.  
Now it is generally recognized that the microlensing searches
towards the Galactic bulge or nearby galaxies
are very important for solutions of a lot of problems
in astronomy and cosmology. As Paczynski (1996) noted, the most important is
the consensus that the microlensing phenomenon has been discovered.  
Now it is impossible to tell which part of the microlensing event
candidates is actually connected with the effect
since probably there are some variable stars among the event candidates,
it could be stellar variability of an unknown 
kind.\footnote{ The microlensing event candidates proposed early
by the EROS collaboration ( \#1 and \#2) and by
the MACHO collaboration (\#2 and \#3) are considered now 
as the evidence of a stellar variability (Paczynski (1996)).}

\begin{enumerate}
\item
Observed light curves are  achromatic and their
shapes are interpreted by simple theoretical expressions very well,
however, there is not complete consent about "very well
interpretation" since even for the event candidate 
MACHO \# 1 the authors of the discovery proposed two fits.
Dominik and Hirshfeld (1994, 1995) 
suggested that the event could be interpreted
very well in the framework of the binary lens model, 
but Gurevich et al. (1996) assumed that
the microlensing event candidate could be caused by
a non-compact microlens.\footnote{Microlensing by non-compact
objects considered also by Zakharov \& Sazhin (1996a,1996b)
and Zakharov (1998,1999,2001a,2001b).}

\item
As expected, binary lenses have been detected
and the observed rate of the events correspond to expected value.

\item
As expected, the parallax effect has been detected.

\item
Since the observed optical depth
is essentially greater than the estimated value,
the independent confirmation of the Galactic bar existence was done.

\item
Using photometric observations of the caustic-crossing
binary lens microlensing event EROS BLG-2000-5, PLANET collaboration
reported about the first microlens mass determination, namely
the masses of these components are 0.35 $M_\odot$ and 0.262 $M_\odot$
and the lens lies within 2.6 kpc of the Sun (An et al. (2002)).

\item
Bennett et al. (2002) discovered gravitational microlensing events due to 
stellar mass black holes. The lenses for events 
MACHO-96-BLG-5
and 
MACHO-96-BLG-6 are the most massive, with mass estimates
$M/M_\odot=6^{+10}_{-3}$
and
$M/M_\odot=6^{+7}_{-3}$, respectively.

\end{enumerate}

Now the following results are generally accepted:

\begin{enumerate}
\item
The optical depth towards the Galactic bulge
is equal to $ \sim 3 \times 10^{-6} $, so it is
larger than the estimated value (Alcock et al. (2000a)).

\item

Analysis of 5.7 years of photometry on 11.9 million stars
in LMC by MACHO collaboration reveals 13 -- 17 microlensing
events (Alcock et al. 2000b).
The optical depth towards the LMC
is equal to $\tau(2 < \hat{t} < 400 {\rm~ days}) 
=1.2^{+0.4}_{-0.3} \times 10^{-7} $, so, it is
smaller than the estimated  value.
The maximum likelihood analysis gives a MACHO halo fraction
f=0.2. 
Alcock et al. (2000b) gives also estimates of
the following probabilities 
$P (0.08 < f <0.5)=0.95$
and $P(f=1) < 0.05.$
The most likely MACHO mass $M \in [0.15, 0.9] M_\odot$,
depending on the halo model and total mass in MACHOs out
50 kpc is found to be $9^{+4}_{-3} \times 10^{10} M_\odot$
EROS collaboration gives a consistent conclusion,
namely,  Lasserre et al. (2000); Lasserre  (2001)
estimate the following probability
$P (M \in [10^{-7}, 1]M_\odot~\&~f>0.4) <~ 0.05$.

%\item
%A lot of new interesting scientific results could be
%extracted from the giant data base which is collected
%during microlensing searches, thereby, as Schneider wrote,
%microlensing searches are "eldorado" for experts in stellar
%variability. New kinds of stellar variability were found already
%using microlensing observations, but probably the data base
%contain other interesting information and have the great
%scientific significance.

\end{enumerate}

However there are different suggestions
(which are not contradicted to the observational data)
about the following issues (Paczynski (1996)):

{\it What is the location of objects which dominate microlensing
observed towards the Galactic bulge?}

{\it Where are the most microlenses for searches towards LMC?}
The microlenses may be in the Galactic disk, Galactic halo,
the LMC halo or in the LMC itself.
{\it Are the microlenses stellar mass objects or are they
substellar brown dwarfs?}

{\it What fraction of microlensing events is caused by
 binary lenses?}

{\it What fraction of microlensing events is connected
with binary sources?}

Paczynski (1996) suggested that we shall have definite answers for
some presented issues after some years and since the optical depth
towards the Galactic bulge is essentially greater than
the optical depth towards the LMC,
we shall have more information about the lens distribution towards
the Galactic bulge, however,
probably, some problems in theoretical interpretation
will appear after detections of new microlensing event candidates.

%\subsection{Conclusions}

The main result of the microlensing searches is that
the effect predicted theoretically has been confirmed.
This is one of the most important astronomical discoveries.

When new observational data would be collected
and the processing methods would be perfected, probably
some microlensing event candidates lost their status, but perhaps
new microlensing event candidates would be extracted among
analysed observational data.
So, the general conclusion may be done.  
The very important astronomical phenomenon was discovered,
but some quantitative parameters of microlensing will be 
specified in future.
However, the problem about 80\% of DM in the halo of our Galaxy is still 
open (10 years ago people believe that microlensing could give
an answer for this problem).

%\section{Bibliographic remarks}

%\subsection{Acknowledgements}

I thank Dr. L. Popovi\'c for his assistance in attending 
the  XIII National Conference of Yugoslav Astronomers.

This research was  supported in part by 
the Russian Foundation for Basic Research (grant \# 00-02-16108).

\references

 Alcock, C., et al.: 1993,
%Possible gravitational microlensing of star
%in Large Magellanic Cloud. 
\journal{ Nat}, \vol{ 365}, 621.
 
 Alcock, C., et al.: 2000a,
\journal{Astroph. J.}, \vol{541}, 734;
\journal{Preprint} astro-ph/0002510.

 Alcock, C., et al.: 2000b,
\journal{Astroph. J.}, \vol{542}, 281;
\journal{Preprint} astro-ph/0001272.

 An, J.H., et al.: 2002,
\journal{Astroph. J.}, \vol{572}, 521;
\journal{Preprint} astro-ph/0110095.

 Aubourg. E., et al.: 1993,
%Evidence for gravitational microlensing by
%dark objects in Galactic halo. 
\journal{ Nat}, \vol{ 365}, 623.

 Bennett, D.C., et al.: 2002,
\journal{Astroph. J.}, \vol{579}, 639;
\journal{Preprint} astro-ph/0109467.

Bogdanov, M. B., Cherepashchuk, A. M.: 1995,
\journal{ Pis'ma v Astron. Zhurn.}, \vol{ 21},  570.

Byalko, A. V.: 1969, \journal{Astron. Zhurn.}, 
\vol{ 46},  998.

 Dominik, M., Hirshfeld, A. C.: 1994,
\journal{ Astron. and  Astrophys.}, \vol{ 289}, L31.

 Dominik, M., Hirshfeld, A. C.: 1995, \journal{ Preprint}
DO-TH 95/19, Dortmund.

Griest, K.: 1991, \journal{ Astrophys. J.}, \vol{ 366},  412.

Gurevich, A. V., Zybin K. P., Sirota V. A.: 1996, \journal{ Phys. Lett. A} 
\vol{ 214}, 232. 

%\bibitem{lan88}
Landau, L. D., Lifshitz, E. M.: 1975, \journal{The Classical Theory
of Fields}, Pergamon Press, Oxford.

 Lasserre, T., et al.: 2000,
\journal{Astron \& Astroph. }, \vol{355}, L39;
\journal{Preprint} astro-ph/0002253.

 Lasserre, T.:  2001,  
\journal{Dark Matter in Astro- and Particle Physics with Gravitational Lensing, 
ed. H.V.~Klapdor-Kleingrothaus}, Proc. of the
Intern. Conf. DARK-2000, Springer, p.~342.

%\bibitem{mol75}
M\"oller, C.: 1972, \journal{The Theory of Relativity}
Oxford, Clarendon Press.
%[Translated into  Russian (Moscow, Atomizdat, 1975)].

Moniez, M. : 2001, \journal{Cosmological Physics with Gravitational Lensing, 
eds. J. Tr\^an Thanh V\^an, Y.~Mellier \& M.~Moniez}, Proc. of the
XXVth Rencontres de Moriond, EDP Sciences, p.~3.  

 Paczynski, B.:  1986, \journal{ Astrophys. J.}, \vol{ 304}, 1.

 Paczynski, B.:  1996, \journal{ Ann. Rev. Astron \& Astrophys.}, \vol{ 34}, 
419.

 Roulet, E., Mollerach, S.: 1997, \journal{Phys. Rep.}
\vol{279}, 2.

Zakharov, A. F.: 1997, \journal{Gravitatsionnie linzi i microlinzi} 
\journal{(Gravitational Lenses and Microlenses)} 
Janus-K, Moscow.

Zakharov, A. F.: 1998,  
\journal{Phys. Lett. A}, \vol{250}, 67.

Zakharov, A. F.: 1999,  
\journal{Astron. Rep.}, \vol{43}, 325.

Zakharov, A. F.: 2001a,  
\journal{Dark Matter in Astro- and Particle Physics with Gravitational Lensing, 
ed. H.V.~Klapdor-Kleingrothaus}, Proc. of the
Intern. Conf. DARK-2000, Springer, p.~364.  

Zakharov, A. F.: 2001b,  
\journal{Cosmological Physics with Gravitational Lensing, 
eds. J. Tr\^an Thanh V\^an, Y.~Mellier \& M.~Moniez}, Proc. of the
XXVth Rencontres de Moriond, EDP Sciences, p.~57.

Zakharov, A. F.,  Sazhin, M.V.: 1996a,  
\journal{Pis'ma Zh. Eksp. Teor. Fiz. }, \vol{ 63}, 894
[\journal{JETP Letters}, \vol{63}, 937].

Zakharov, A. F.,  Sazhin, M.V.: 1996b,  
\journal{Zh. Eksp. Teor. Fiz. } \vol{110}, 1921
[\journal{JETP} \vol{83}, 1057].

Zakharov, A. F.,  Sazhin, M.V.: 1998,  
\journal{Usp. Fiz. Nauk}, \vol{ 168}, 1041
[\journal{Phys. Usp.}, \vol{41}, 945 ].

%Hewitt, A., Burbrdge, G. : 1989, \journal{Astrophys. J. Suppl. Series}, \vol{75}, 297.

%Mediavilla, E., Insertis, F. M. : 1989, \journal{Astron. Astrophys.} \vol{214}, 79. 

%Netzer, H. : 1990, \journal{Active Galactic Nuclei, eds. R. D. Blandford, H. Netzer \& L. Woltjer}, Saas-Fee Advanced Course 20, Berlin: Springer -- Verlag.  

%Osterbrock, D. E. : 1989, \journal{Astrophysics of Gaseous Nebulae and Active Galactic Nuclei}, Mill Valley, California. 
\endreferences

\end{document}